\documentclass[prd,preprintnumbers,superscriptaddress,nofootinbib,twocolumn]{revtex4-2}
\usepackage{epsfig}
\usepackage{amsmath}
\usepackage{amsfonts}
\usepackage{float}
\usepackage{amssymb}
\usepackage{slashed}
\usepackage{color}
\usepackage{xcolor}
\usepackage{pbox}
\usepackage{subfigure}
\usepackage[colorlinks,citecolor=black, linkcolor = black, urlcolor = black]{hyperref}
\usepackage{tabularx}
\usepackage{titlesec}
\usepackage{scrextend}
\usepackage[version=3]{mhchem}

\def\lsim{\lesssim}
\def\gsim{\gtrsim}

\def\be{\begin{eqnarray}}
\def\ee{\end{eqnarray}}

\def\nn{\nonumber}

\def\D{\text{D}}

\def\bal#1\eal{\begin{align}#1\end{align}}
\def\beq{\begin{equation}}
\def\eeq{\end{equation}}
\def\l{\left}
\def\r{\right}
\def\GeV{{\rm GeV}}
\def\MeV{{\rm MeV}}
\def\keV{{\rm keV}}

\def\H{\text{H}}
\def\T{\text{T}}

\def\tnu{{\nu_{\rm nt}}}
\def\tanu{{\nu_{\rm nt, \alpha}}}
\def\bnu{{\nu_{\rm bg}}}

\def\tnubar{{\bar \nu_{\rm nt}}}
\def\tanubar{{\bar\nu_{\rm nt, \alpha}}}
\def\bnubar{{\bar \nu_{\rm bg}}}

\def\nud{{\nu {\rm d}}}

\def\Tgamma{{T}} 
\def\Tgammastar{{T_{*}}} 

\begin{document}

\title{Constraining MeV to 10 GeV majoron by Big Bang Nucleosynthesis}
\author{Sanghyeon Chang}
\email{schang@ibs.re.kr}
\affiliation{Particle Theory  and Cosmology Group (PTC), 
Center for Theoretical Physics of the Universe (CTPU), \\
Institute for Basic Science, Daejeon 34126, Republic of Korea}

\author{Sougata Ganguly}
\email{sganguly@cnu.ac.kr}
\affiliation{Department of Physics and Institute of Quantum Systems (IQS), \\ 
Chungnam National University, Daejeon 34134, Republic of Korea}

\author{Tae Hyun Jung}
\email{thjung0720@gmail.com}
\affiliation{Particle Theory  and Cosmology Group (PTC),
Center for Theoretical Physics of the Universe (CTPU), \\
Institute for Basic Science, Daejeon 34126, Republic of Korea}

\author{Tae-Sun Park}
\email{tspark@ibs.re.kr}
\affiliation{Center for Exotic Nuclear Studies (CENS), \\
Institute for Basic Science, Daejeon 34126, Republic of Korea}

\author{Chang Sub Shin}
\email{csshin@cnu.ac.kr}
\affiliation{Department of Physics and Institute of Quantum Systems (IQS), \\
Chungnam National University, Daejeon 34134, Republic of Korea}
\affiliation{Particle Theory  and Cosmology Group (PTC), 
Center for Theoretical Physics of the Universe (CTPU), \\
Institute for Basic Science, Daejeon 34126, Republic of Korea}
\affiliation{Korea Institute for Advanced Study, Seoul 02455, South Korea}

\preprint{CTPU-PTC-23-56}

\begin{abstract}
We estimate the Big Bang nucleosynthesis (BBN) constraint on the majoron
in the mass range between $1\,{\rm MeV}$ to $10\,{\rm GeV}$ which dominantly decays into the standard model neutrinos. 
When the majoron lifetime is shorter than $1\,{\rm sec}$, 
the injected neutrinos mainly heat up background plasma, which alters the relation between photon temperature and background neutrino temperature.
For a lifetime longer than $1\,{\rm sec}$, 
most of the injected neutrinos directly contribute to the protons-to-neutrons conversion.
In both cases, 
 deuterium and helium abundances are enhanced, while the constraint from the deuterium is stronger than that from the helium.
 $\ce{^7Li}$ abundance gets decreased as a consequence of additional neutrons, but the parameter range that fits the observed $\ce{^7Li}$ abundance is excluded by the deuterium constraint.
We also estimate other cosmological constraints and compare them with the BBN bound.
\end{abstract}

\maketitle

\section{Introduction}
The analysis of the Big Bang nucleosynthesis (BBN) has successfully predicted primordial abundances of light elements such as $\ce{^4He}$, 
$\D$, and $\ce{^3He}$ 
(see Ref.\,\cite{Pitrou:2018cgg, Fields:2019pfx} for a review).
The primordial $\ce{^4He}$ and $\D$ abundances are precisely measured by a few percent level accuracies\,\cite{ParticleDataGroup:2022pth},
and they agree well with the standard BBN (SBBN) prediction with the baryon asymmetry input $\eta_b\equiv n_b/n_\gamma =6.1\times 10^{-10}$ that is obtained by fitting the cosmic microwave background (CMB) data\,\cite{Planck:2018vyg}.
The $\ce{^3He}$ abundance was recently measured by Ref.\,\cite{Cooke:2022cvb} within an agreement with SBBN although there is a theoretical uncertainty coming from models of the galactic chemical evolution.
On the other hand, the long-standing problem of the observed $\ce{^7Li}$ abundance being smaller than the SBBN prediction still remains unsolved\,\cite{Asplund:2005yt,Aoki:2009ce,Sbordone:2010zi, Melendez:2010kw}.

The success of SBBN analysis has provided strong constraints on new particles that (partially) decay to standard model (SM) particles around the BBN era.
Even when a new particle dominantly decays to neutrinos which have the weakest coupling to nucleons, the BBN analysis gives meaningful constraints\,\cite{Scherrer:1984,Chang:1993yp,Kanzaki:2007pd,Pospelov:2010cw,Fradette:2017sdd,Blinov:2019gcj,Sabti:2019mhn}.

In this paper, we estimate the BBN constraint on a (pseudo-)scalar particle that decays to neutrinos.
Motivated by the majoron model\,\cite{Chikashige:1980ui,Gelmini:1980re},
we consider a model where the majoron $J$ interacts with neutrinos as
\bal
{\cal L}_{\rm int} = - \frac{g_{\alpha \beta}}{2} J \, \nu^T_\alpha \sigma_2 \nu_\beta + {\rm h.c.},
\eal
where $\nu_\alpha$ is the SM neutrino with flavor $\alpha = e,\,\mu,\,\tau$.
For simplicity, we assume the flavor universality, i.e. $g_{\alpha \beta} = g \delta_{\alpha\beta}$.
We expect that a dedicated analysis for the realistic majoron model ($g_{\alpha \beta} \simeq m_{\nu \alpha \beta}/f_J$ for the $B-L$ symmetry breaking scale $f_J$ and the mass matrix of SM neutrinos $m_{\nu \alpha \beta}$ in the flavor eigen-basis) would not be much different from our results because the individual elements of the Pontecorvo–Maki–Nakagawa–Sakata matrix (PMNS matrix) are all of order one\,\cite{ParticleDataGroup:2022pth}.

The BBN constraint on the majoron model was estimated in Refs.\,\cite{Chang:1993yp,Blinov:2019gcj,Sabti:2019mhn} based on the change in the expansion rate; the enhanced expansion rate makes neutron-proton freeze-out 
earlier, which leads to an increase in the neutron-to-proton ratio.
They focused on the range of majoron mass and coupling, $m_J \lsim 10\,\MeV$ and  $g\gsim 10^{-10}$, respectively, so that majorons are in the thermal bath and contribute to the relativistic degrees of freedom during the BBN era (see also Ref.\,\cite{Berryman:2022hds} for a comparison to other constraints).

Here, we focus on scenarios where the majoron has already been thermally decoupled before the BBN era.
In this case, a long lifetime of the majoron can cause nontrivial effects on the BBN. 
The lifetime of $J$ is given by
\bal
\tau_J &\equiv \Gamma_J^{-1} = (2\times 3 \, \Gamma_{J \to \nu_\alpha \nu_\alpha})^{-1}
=\frac{16\pi}{3g^2 m_J} \nonumber\\
&\simeq 0.11\,\sec \, \left(\frac{10^{-11}}{g} \right)^{2} \left( \frac{\GeV}{m_J} \right)
\eal
where $\Gamma_{J\to \nu_\alpha \nu_\alpha}=\Gamma_{J\to \bar\nu_\alpha\bar\nu_\alpha} = g^2m_J/32\pi$ is the partial decay width of individual $J\to \nu_\alpha \nu_\alpha$, and $J\to \bar\nu_\alpha\bar\nu_\alpha$. 
We distinguish $\nu$ and $\bar \nu$ by the helicity (or chirality).
Since the BBN process starts around $t_\nud \sim 0.1\,\sec$ when the background neutrinos are decoupled, our analysis is relevant for $g< 10^{-11} (\GeV/m_J)^{1/2}$.

The majoron mass range in our analysis is restricted as
$1\, {\rm MeV} \le m_J \le 10\,{\rm GeV}$ 
for the following reasons. 
Because the neutrino decoupling temperature is about $2\,\MeV$, if the majoron is lighter than $\MeV$, the injected neutrinos do not modify the BBN process except for contributing to an additional source of energy density.
In this case, constraint from the change in the effective number of neutrino species ($\Delta N_{\rm eff}$)
from the CMB analysis is stronger than the BBN bound.
For majorons heavier than $10\,\GeV$, the energy of injected neutrinos is so high that various channels including muons, pions, etc., must be involved. 
We avoid such complexity in our analysis by restricting the mass range of majoron (see, e.g. Ref.\,\cite{Kanzaki:2007pd} for the case of neutrino injection energy higher than $O(100)\,\GeV$). 

The majoron initial abundance strongly depends on the reheating temperature of the universe and the underlying UV model of the majoron.
For instance, if the universe undergoes the $B-L$ cosmic phase transition from which the majorons are produced (see, e.g. Ref.\,\cite{Huang:2022vkf,Dasgupta:2022isg,Chun:2023ezg} for a relevant leptogenesis scenario), 
the majoron yield $Y_J=n_J/s$ is frozen at high temperature and its value at the beginning of the BBN procedure
$Y_J^{(0)}= n_J/s$ at $T=10\,\MeV$ is given by $0.28/g_{*s}(T_{B-L})$,
where $T_{B-L}$ is the $B-L$ phase transition temperature, $n_J$ is the majoron number density, $s$ is the entropy density, and $g_{*s}$ is the effective degrees of freedom for the entropy density.
On the other hand, if the $B-L$ symmetry had never been restored, the majorons could be produced through the freeze-in process.
To avoid too much model-dependent discussion, we treat $Y_J^{(0)}$ as a free parameter and present our constraints in terms of upper bound on $Y_J^{(0)}$ and $\tau_J$ for different $m_J$.
We also provide exclusion plots projected in the $(m_J, g)$ plane for several choices of $Y_J^{(0)}$.

The rest of the article is organized as follows. In Sec.\,\ref{sec:modifications}, we discuss the modifications of the BBN processes and Sec.\,\ref{sec:results} is dedicated
to the numerical results. Finally, we conclude in Sec.\,\ref{sec:conclu}. The relevant expression for the momentum distribution of nonthermal neutrinos
and their cross sections with $n$, $p$, $\D$, $\ce{^4 He}$ are given in Appendix \ref{App:scattering} and \ref{App:cross_section}
respectively. The reaction rates for $n \leftrightarrow p$ conversion processes are given in Appendix \ref{app:n2p_conversion}.

\section{Modification of the BBN process}
\label{sec:modifications}
The late-time injection of neutrinos can modify the BBN scenario 
in the following ways:
\begin{itemize}
\item[1.]{Injected neutrinos directly contribute to nuclear reactions via the weak interaction.}
\item[2.]{Background neutrino ($\bnu$, $\bnubar$) and visible plasma ($e\gamma B$) are heated differently, modifying the relation between their temperatures.}
\item[3.]{The expansion rate is modified.}
\end{itemize}
In order to correctly take into account these effects, the evolution of injected neutrino distribution should be consistently treated.

We simplify the analysis by assuming that a single scattering or annihilation of an injected neutrino sufficiently reduces its initial energy and makes it merge into the background plasma, which means that the energy of an injected neutrino is redistributed to the background particles by one scattering or annihilation.
As a result, our simplified distribution contains fewer neutrinos in the intermediate energy range compared to the actual distribution of neutrinos.
This leads to an underestimation of the interaction rate with nuclei induced by injected neutrinos because of the short-distance property of the weak interaction and provides a conservative estimation of the BBN constraint.

Our estimation is not too conservative because our assumption still gives an approximately correct distribution in high-energy regions, whose contribution to the BBN modification is most dominant.
Therefore, we do not expect a significant difference to be made by a more realistic analysis which may be done by solving the full Boltzmann equation of the whole neutrinos without separating the background neutrinos and the energetic neutrinos.

In the following subsections, we explain how we estimate the distribution function of high-energy neutrinos, the heating effects on $e\gamma B$ and $\bnu$ sectors, the modified Hubble rate, and $\Delta N_{\rm eff}$.
Subsequently, we describe the effect of these quantities on BBN.

\subsection{Distribution function of energetic neutrinos}
\label{Sec:f_tnu}

First, let us focus on the distribution function of nonthermally produced energetic neutrinos with a flavor $\alpha=e,\mu,\tau$ denoted by $\tanu$ ($\tanubar$ for anti-neutrino)\,\footnote{
In our mass and temperature range ($1\,\MeV \leq m_J \leq 10\,\GeV$ and $T\lsim 1\,\MeV$), only the electron flavor of injected neutrinos can induce nuclear reactions.
}. 
The Boltzmann equation for the distribution 
$f_{\tanu}(t, p)$ of $\tanu$ can be written as
\beq
\frac{\partial f_{\nu_{\rm nt, \alpha} }}
{\partial t}
- H p \frac{\partial f_{\tanu}}
{\partial p}
=\sum_i {\cal C}_{\alpha i} 
\label{Eq:BE_original}
\eeq
with the Hubble rate $H$, the magnitude of the majoron momentum $p=|\vec p|$, and collision terms ${\cal C}_i$.
The source term of $J\to {\nu_{\rm nt, \alpha} } {\nu_{\rm nt, \alpha} }$ 
($J\to {\bar{\nu}_{\rm nt,\alpha}} {\bar{\nu}_{\rm nt,\alpha}}$ for $f_{\bar{\nu}_{\rm nt,\alpha}}$) can be written as
\bal
{\cal C}_{J\to \tanu\tanu} &= 
\frac{1}{E}
\int d\Pi_J d \Pi_{\nu_\alpha} 
|{\cal M}_{J\to \nu_\alpha \nu_\alpha}|^2 f_J  \nonumber \\
& \qquad \qquad 
\times  (2\pi)^4 \delta^{(4)}(P_J - P-P_{\nu_\alpha} ) 
\nn
\\
&= \frac{2\pi^2 \, \Gamma_J n_J}{3 E^2}\,
\delta \left(E  - \dfrac{m_J}{2}\right),
\eal
where $P^\mu=(E, \vec p)$, and  $d\Pi_i = d^3 \vec p_i/((2\pi)^3 2E_i)$ is the phase space integration, we used the total decay width $\Gamma_J= 6\Gamma_{J\to \nu_\alpha\nu_\alpha}$, and neglected Pauli blocking factors.
For a given initial yield of majoron $Y_J^{(0)}$, the majoron number density is evolved as $n_J\simeq Y_J^{(0)} s(T) e^{-\Gamma_J t}$.
Other scattering terms with the background plasma can be written as
\bal
{\cal C}_{\tanu a \to bc} &= - \frac{S}{2E}
\int 
d\Pi_a
d\Pi_b
d\Pi_c
|{\cal M}_{\nu_\alpha a\to bc}|^2 f_a f_\tanu \nn\\
& \qquad\quad
\times (2\pi)^4 \delta^{(4)}(P + P_a - P_b -P_c),  
\label{Eq:C_scattering}
\eal
where $S$ is the symmetry factor.
We do not include processes of ${\cal C}_{bc \to a \tanu}$ as we consider those scattered neutrinos to be a part of the background neutrinos (so we consider all the elastic scattering as $\tanu a \to \bnu a$).
This provides a conservative estimation of the energetic neutrinos as we discussed previously.

Then, using the dimensionless parameters $z=m_e/T$, $\xi=p/T$,  
Eq.~(\ref{Eq:BE_original}) is organized 
as\footnote{
We neglect corrections in the change of variables from $(t,p)$ to $(z,\xi)$ which arise when the temperature crosses the electron threshold.
The error coming from the electron threshold is $O(10)\%$.
}
\bal
\frac{\partial f_\tanu}{\partial z}
= 
    A_\alpha(\xi, z) \, \delta \l(z- \frac{2\xi m_e}{m_J}\r) - B_\alpha(\xi,  z) f_\tanu
 , 
\label{Eq:Boltzmannn_AB}
\eal
where $A_\alpha(\xi, z)$ and $B_\alpha(\xi,z)$ correspond to the source term and the scattering term,
\bal
A_\alpha(\xi, z) &= \frac{16\pi^4 g_{*s} }{135} \frac{ m_e^2  \Gamma_{J} Y_{J}^{(0)}}{ m_J^2 } \,
\frac{ e^{-\Gamma_J/2H(z)}}{\xi z^2 H(z)}
\label{Eq:Aterm}\,,\\
B_\alpha(\xi, z) &=
\frac{7\pi G_F^2 m_e^5}{90H}\frac{\xi}{z^6} \Bigg[
 \zeta_{\alpha 1} \, \theta(T-m_e) \label{Eq:Bterm}
  \\ 
&  + \left(\frac{T_\bnu}{T}\right)^4
\Big(\zeta_{\alpha 2} +  \zeta_{\alpha 3} \, \theta(E\, T_\bnu - m_e^2)\Big) 
\Bigg] ,
\nn
\eal
where $G_F$ is the Fermi constant, and we take $t\simeq 1/2H$ approximation.
The values of constants $\zeta_{\alpha 1}$, $\zeta_{\alpha 2}$, and $\zeta_{\alpha 3}$ for different flavors are summarized in Appendix\,\ref{App:scattering}. 

The solution of Eq.\,\eqref{Eq:Boltzmannn_AB} is given by
\bal
f_\tanu (\xi, z)
&=
A_\alpha\l(\xi, \frac{2\xi m_e}{m_J} \r)
\,
\theta\l(z-\frac{2\xi m_e}{m_J}\r)
\nonumber \\
&\quad
\times
\exp
\l[
    -\int_{\frac{2\xi m_e}{m_J}}^z
     dz'  B_\alpha (\xi, z') 
\r].
\label{Eq:ftnu_general}
\eal
We take $f_\tanubar = f_{\tanu}$ since $A_\alpha$ and $B_\alpha$ terms are the same for $\nu_\alpha$ and $\bar \nu_\alpha$ except for neutrino-baryon interaction rate whose contribution is highly suppressed by the small baryon number density compared to that of photons 
$\eta_b\sim 10^{-9}$.
On the other hand, in the Boltzmann equations for the abundance of light nuclei, the interaction rates between $\tnu$ and baryons are non-negligible compared to other nuclear reaction rates and thus should be included.

\subsection{Heating effects}
The scattering/annihilation of injected neutrinos with the background plasma heats up the standard plasma ($e\gamma B$) as well as the background neutrinos ($\bnu$).
With our assumption of neutrino distribution, we provide a good approximation to estimate the changes in background temperatures of neutrinos $T_\bnu$ and photons $\Tgamma$.
Remind that our analysis provides a conservative estimation of the constraints as we discussed earlier.

The process of nucleosynthesis is completely insensitive to the overall heating prior to neutrino decoupling at $t=t_\nud$ ($T=T_\nud$)
(except for adjusting the baryon asymmetry parameter). 
When the neutrinos are injected before the neutrino decoupling period $t<t_\nud$ ($T>T_\nud$), they get quickly thermalized, and their energy is efficiently redistributed to the background neutrinos and the electromagnetic plasma with a common temperature $T_\nu =T$.
Therefore, we only take into account the residual decays of majorons after the neutrino decoupling.

Then, for $ t\geq t_\nud $ ($ T_\bnu,  \Tgamma
<T_\nud$) we have the Boltzmann equations for the background neutrinos $\bnu$ and electromagnetic plasma with the assumption of the simplified distribution of neutrinos as 
\bal
&\dot\rho_{\tanu} + 4H \rho_{\tanu} \!\!
=  \frac{\rho_J}{3\tau_J} 
-  {\cal W}(\tanu\to \bnu)
\nn \\ 
& \hskip 3.55cm 
-  {\cal W}(\tanu \to e)\,\,,   
\label{Eq:rho_tnu} 
\\
&\dot\rho_{\bnu} + 4 H \rho_{\bnu}  
= \!\! \!\!  \sum_{\alpha=e, \mu, \tau} \!\! {\cal W}(\tanu \to \bnu)
\nonumber\\
& \hskip 2.5cm   +  {\cal W}( e \to \bnu)\,,
\label{Eq:rho_bnu}
\\
&\dot\rho_{e\gamma B} + 3H  (\rho_{e\gamma B} + P_{e\gamma B})
 = \!\!\!\!  \sum_{\alpha=e,\mu,\tau} \!\!  {\cal W}(\tanu \to e) 
\nonumber\\
& \hskip 4.2cm 
- {\cal W}(e\to \bnu)\,\,,
\label{Eq:rho_plasma}
\eal 
where $\rho_{e\gamma B}$ is mostly dominated by relativistic degrees of freedom, so $P_{e \gamma B} \approx \rho_{e\gamma B}/3$.
The majoron energy density evolves as 
\bal
\rho_{J} = m_J  Y_J^{(0)}s(
\Tgamma
) e^{-t/\tau_J}\,\,,
\eal
and the energy transfer functions are given by
\bal
& {\cal W}(\tanu \to \bnu) =  \Gamma(\tanu\to \bnu)\,\rho_{\tanu}\,, \label{Eq:W1}\\
& {\cal W}(\tanu \to e ) \quad =   \Gamma(\tanu\to e)\,\rho_{\tanu} \label{Eq:W2}.
\eal 
Here $\Gamma(\tanu\to \bnu)$ and $\Gamma(\tanu\to e)$ are averaged scattering rates for the energy transfer from the injected nonthermal neutrinos $\tanu$ to the background neutrinos and charged leptons, respectively (see Appendix\,\ref{App:scattering} for their expressions). 
Notice that Eq.~(\ref{Eq:rho_tnu}) is the result of Eq.~(\ref{Eq:Boltzmannn_AB}), and
Eq.~(\ref{Eq:rho_bnu}) and~(\ref{Eq:rho_plasma}) show that
background temperatures $T_\bnu$ and $T$ evolve differently from the SBBN. 
The ${\cal W}( e\to \bnu)$ term which already exists in the SBBN becomes small at $t>t_\nud$, but non-negligible.

We provide analytic approximations of the temperature changes by the leading order in the $\rho_J/T^4$ expansion.
Taking 
\bal \Gamma_\alpha \equiv \Gamma(\tanu\to \bnu) +\Gamma(\tanu\to e)
\eal 
as the averaged  total rate of reducing $\rho_\tanu$, the solution of $\rho_\tanu$ is given by 
\bal \label{Eq:rho_tnu2}
& \rho_\tanu (t) \simeq  
 \frac{1}{3} m_J Y_J^{(0)} s(T) 
 \nn\\
   & \quad\times \int_{t_\nud}^t  \frac{dt'}{\tau_J} \,
 \left( \frac{s(T)}{s(T')}\right)^{\hskip-0.1cm \frac{1}{3}}  \exp\left[ - \frac{t'}{\tau_J} - \int_{t'}^{t} dt''\,
 \Gamma_{\alpha}(t'')\right],
\eal
and the heating contributions to the background densities are
\bal
\Delta \rho_\bnu  &  \simeq    \hskip -0.1cm \sum_\alpha  
\int_{t_\nud}^t  \hskip -0.15cm dt'   \left(\frac{s(T)}{s(T')}\right)^{\hskip-0.1cm \frac{4}{3}} \! \Gamma(\nu_{{\rm nt},\alpha} \to \bnu) \, \rho_{\tanu}(t'), 
\\
\Delta \rho_{e\gamma B} & \simeq  \hskip -0.1cm  \sum_\alpha  
\int_{t_\nud}^t \hskip -0.15cm dt' \left(\frac{s(T)}{s(T')}\right)^{\hskip-0.1cm \frac{4}{3}} \! \Gamma(\nu_{{\rm nt},\alpha} \to e)\, \rho_{\tanu}(t') .
\eal
Here, we neglected the entropy increase effect on $a(t')/a(t) = (s(T)/s(T'))^{1/3}$ due to the majoron decay which is the next leading order in $Y_J^{(0)}$ expansion.

If $\tau_J \lesssim  t_\nud$, the dominant contribution is made around $t\sim t_\nud$, and each contribution at that time is estimated as
\bal
\frac{\Delta\rho_{\bnu}}{\rho_{e\gamma B}} &\simeq 
\frac{86}{99}\sum_\alpha  x_\alpha\kappa_\alpha  
\frac{m_J Y_J^{(0)}}{  T_{\nud}} 
e^{- t_\nud/\tau_J} , 
\label{Eq:Deltarhobnu_bf_tnud}
\\
\frac{\Delta\rho_{e\gamma B}}{\rho_{e\gamma B}} &\simeq   \frac{86}{99}\sum_\alpha  x_\alpha(1-\kappa_\alpha)  \frac{m_J Y_J^{(0)}}{  T_{\nud}} 
e^{- t_\nud/\tau_J} ,
\label{Eq:DeltarhoegB_bf_tnud}
\eal
where the prefactor comes from $s(T_{\nud})/3\rho_{e\gamma B} = 86/99 T_\nud$.
For simplicity of the formulae, we have introduced time-dependent efficiency factors $x_\alpha$ and $\kappa_\alpha$ as
\bal
x_\alpha= 1-e^{-\Gamma_\alpha t }, \quad
\kappa_\alpha = \frac{\Gamma(\tanu \to \bnu)}{\Gamma_\alpha}
\label{Eq:eff_alpha}
\eal
which should be evaluated at $t=t_\nud$ in Eq.\,\eqref{Eq:Deltarhobnu_bf_tnud} and \eqref{Eq:DeltarhoegB_bf_tnud}.

For the case of $\tau_J \gtrsim t_\nud$, we should in principle take into account continuously injected nonthermal neutrinos from the decay of majorons.
After neutrino decoupling ($t_\nud \lesssim t \lesssim \tau_J$), the energy density of the injected neutrinos relative to the background radiation gradually increases as $\propto (m_J/T)\cdot (\Gamma_J t)$.  
Together with $\Gamma_\alpha \propto T^4$ (see Appendix\,\ref{App:scattering} for explicit expressions), we find that the largest heating contribution occurs when the age of the universe approaches majoron lifetime, i.e. at $t\sim \tau_J$, although the scattering rate $\Gamma_\alpha $ can be quite suppressed. 
Therefore, if $\tau_J > t_\nud$, the additional energy densities at
 $t\sim \tau_J$  are estimated as 
\bal
\frac{\Delta\rho_{\bnu}}{\rho_{e\gamma B}} &\simeq \frac{86}{99}\sum_\alpha x_\alpha\kappa_\alpha  \frac{m_J Y_J^{(0)}}{  T_{\rm decay}} 
e^{- t_\nud/\tau_J} ,  \\
\frac{\Delta\rho_{e\gamma B}}{\rho_{e\gamma B}} &\simeq \frac{86}{99}\sum_\alpha x_\alpha(1-\kappa_\alpha)  \frac{m_J Y_J^{(0)}}{  T_{\rm decay}} 
e^{- t_\nud/\tau_J} , 
\eal
where $T_{\rm decay}$ is the photon temperature at majoron decay ($t=\tau_J$) and $x_\alpha$, $\kappa_\alpha$ are evaluated at $t=\tau_J$.
The relevant quantity for the BBN is the ratio between the background neutrino energy density and that of the plasma (photon).
From the previous discussions, the deviation of the ratio compared to that for the Standard BBN (SBBN) is obtained as  
\bal
\dfrac{\Delta T_\bnu}{T_\bnu} &= \left[\dfrac{1 + \Delta \rho_{\bnu}/\rho_\bnu}{1 + \Delta \rho_{e \gamma B}/\rho_{e \gamma B}}\right]^{1/4} - 1 \nn\\
& = \left[\dfrac{1 + \sum_\alpha \frac{c_* m_J Y_J^{(0)}x_\alpha \kappa_\alpha e^{-t_{\nu_d}/\tau_J}}{3 T_*}}
{1 + \sum_\alpha \frac{2.61 m_J Y_J^{(0)}x_\alpha (1-\kappa_\alpha) e^{-t_{\nu_d}/\tau_J}}{3 T_*}}\right]^{1/4}
-1
\label{Eq:ratio_rho}
\eal
for given plasma temperatures $T_*\equiv \min(T_\nud, T_{\rm decay})$. 
$c_*$ is estimated as $2.73\,(3.83)$ for $T_* > m_e$ ($T_* < m_e $).
 
\subsection{Corrections to the expansion rate and $\Delta N_{\rm eff}$}

When the universe expands dominantly by the radiation energy density as
$\rho_{\rm rad}\simeq 3H^2M_P^2$ where $M_P = 2.43 \times 10^{18}\,\GeV$ is
the reduced Planck mass, the effective number of relativistic neutrino species after $e^+e^-$ annihilation, $N_{\rm eff}$, is defined as 
\bal 
N_{\rm eff} =\frac{8}{7}\left(\frac{11}{4}\right)^\frac{4}{3} \left(\frac{\rho_{\rm rad}- \rho_{e\gamma B}}{ \rho_{e\gamma B}}\right).
\eal
In our study, 
\bal 
\rho_{\rm rad}  = \rho_{e\gamma B}+  \rho_{\bnu} + \rho_{\tnu}.
\eal
The additional effective number of relativistic degrees of freedom is given by
\bal
\Delta N_{\rm eff} = 3 \left[\left(1 + \dfrac{\Delta T_\bnu}{T_\bnu}\right)^4  -1\right] + 
\frac{8}{7}\left(\frac{11}{4}\right)^\frac{4}{3}\sum_\alpha \frac{\rho_{\tanu}}{\rho_{e\gamma B}}
\label{Eq:Neff}
\eal
with the information of Eq.~(\ref{Eq:rho_tnu2}) and~(\ref{Eq:ratio_rho}).   
In the calculation of the Hubble rate, we also include the contribution of majoron energy density as $\rho_{\rm tot} =\rho_{e\gamma B} + \rho_{\bnu} + \rho_{\tnu} +\rho_J
=3 H^2 M_P^2$.   

\subsection{Implementation to the BBN code}

Now, let us consider the impact of $\tnu$, $\tnubar$ and $\Delta T_\bnu$ on the Boltzmann equations of nuclei,
\bal
 \frac{d X_A}{dt} = 
 \left. \frac{d X_A}{dt}\right|_{\text{SBBN}} \!\!
 - \sum_B \Big[ \delta \Gamma_{A\to B} X_A - \delta \Gamma_{B\to A} X_B\Big],
\eal
where $X_A\equiv n_A/n_b$ with $n_b$ the baryon number density, 
$\left. (dX_A/dt)\right|_{\text{SBBN}}$
stands for the terms existing in the SBBN, and $A, \, B =p, \, n, \, \D, \, \T, \, \ce{^3He},\, \cdots$ are indices for the light elements.
The coefficient $\delta \Gamma_{A\to B}$ is given as 
\bal
\delta \Gamma_{A\to B} &=\dfrac{1}{2 \pi^2} \int dE_{\tnu}\, E_{\tnu}^2 \,f_{\tnu}\,(\sigma v)_{\tnu A \to B e^-} \nn \\
&\quad+\dfrac{1}{2 \pi^2} \int dE_{\tnubar}\, E_{\tnubar}^2 \,f_{\tnubar}\,(\sigma v)_{\tnubar A \to B e^+}  \nn \\
& \quad+ (\Gamma^\prime_{A\to B} - \Gamma^{(\rm SBBN)}_{A \to B})\,,
\label{Eq:delta_Gamma}
\eal
where $\Gamma_{A\to B}^{\rm (SBBN)}$ is the reaction rate of $\nu A\to B e^-$ or $\bar \nu  A\to B e^+$ that exists in the standard BBN.
$\delta \Gamma_{B\to A}$ can be obtained by replacing $A \leftrightarrow B$.
These corrections are only included for $A=p, \, n,\,\D$ and $\ce{^4He}$. In Appendix \ref{App:cross_section}, we summarize our treatment. 
The last term in Eq.\,\eqref{Eq:delta_Gamma} accounts for the increase of background neutrino temperature, 
which is relevant before the neutron freeze-out. Therefore we only include the last term for $A,B=n$ or $p$.
The form of $\Gamma^\prime_{A\to B}$ for $A, \,B = n, \,p$ is given by
\bal
&\Gamma^\prime_{n\to p} = \tau_n^{-1} + x_{np} (\Gamma^{(\rm SBBN)}_{n\to p} - \tau_n^{-1}) \,,\nn\\
&\Gamma^\prime_{p\to n} = x_{pn} \Gamma^{(\rm SBBN)}_{p\to n} \,.
\eal
Here $\tau_n= 879.4\,\rm sec$ is the neutron lifetime \cite{ParticleDataGroup:2022pth}, and the explicit form of $x_{np}$ and $x_{pn}$  can be found in Appendix \ref{app:n2p_conversion}.

We take into account the modified evolution of $\rho_{e\gamma B}$ which is given by Eq~\eqref{Eq:rho_plasma}.
This can be effectively done by including the correction of ${\cal N}(z)$, the entropy transfer from the incomplete neutrino decoupling in the SBBN\,\cite{Mangano:2001iu,Mangano:2005cc,Mangano:2006ar} as follows;
\bal
&\dot \rho_{e\gamma B} + 3H(\rho_{e\gamma B} + P_{e\gamma B})  
= -T^4 H(T) ({\cal N}(z) +\Delta {\cal N}(z))
\label{Eq:calN}
\eal
where  $\Delta {\cal N}(z)=-\sum_{\alpha}{\cal W}(\tanu \to e )/T^4 H$.

This also causes a dilution of
the baryon asymmetry parameter $\eta_b \equiv n_b/n_\gamma$;
\bal
\frac{\eta_{b,{\rm ini}}}{\eta_{b, {\rm fin}}} =  2.73 - \dfrac{45}{2\pi^2 {g_*s}(T_f)} 
\int_{T_{\nud}}^{T_f} \dfrac{\cal W}{H(T) T^2 T_{\bnu}^3}dT
.
\label{Eq:etaB_dilute}
\eal
Here we fix $\eta_{b,{\rm fin}}=6.1\times 10^{-10}$ and the final temperature in our code ($T_f$) is taken to be $5\,{\keV}$.

In summary, to obtain the final abundance, we implement the Eq.\,\eqref{Eq:delta_Gamma}, \eqref{Eq:calN} and \eqref{Eq:etaB_dilute} as well as the modified Hubble rate corresponding to Eq.\,\eqref{Eq:Neff} to the public code \texttt{PArthENOPE}\,\cite{Pisanti:2007hk,Consiglio:2017pot,Gariazzo:2021iiu} which uses nuclear reaction rates summarized in Ref.\,\cite{Serpico:2004gx}.

\begin{figure}
    \centering
   \includegraphics[width = 0.48\textwidth]{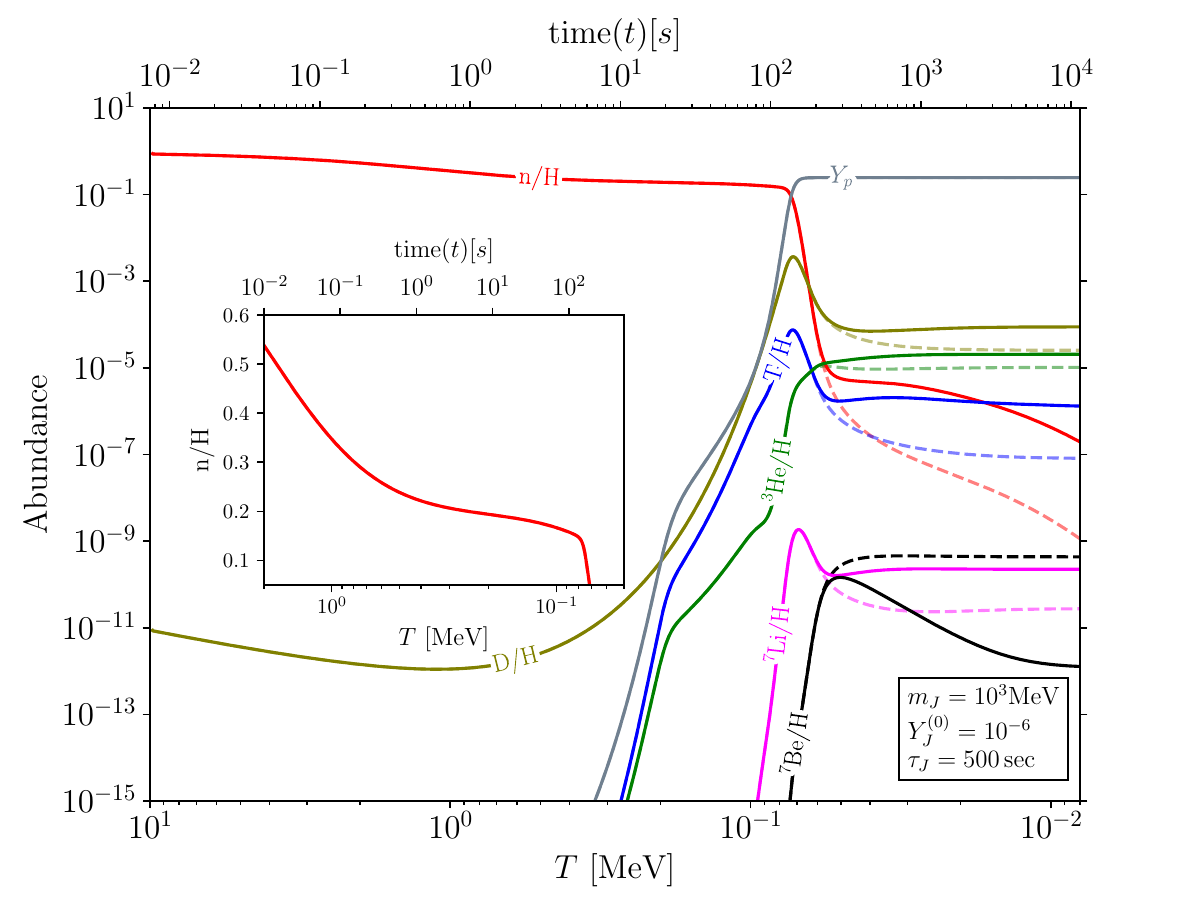}
    \includegraphics[width = 0.48\textwidth]{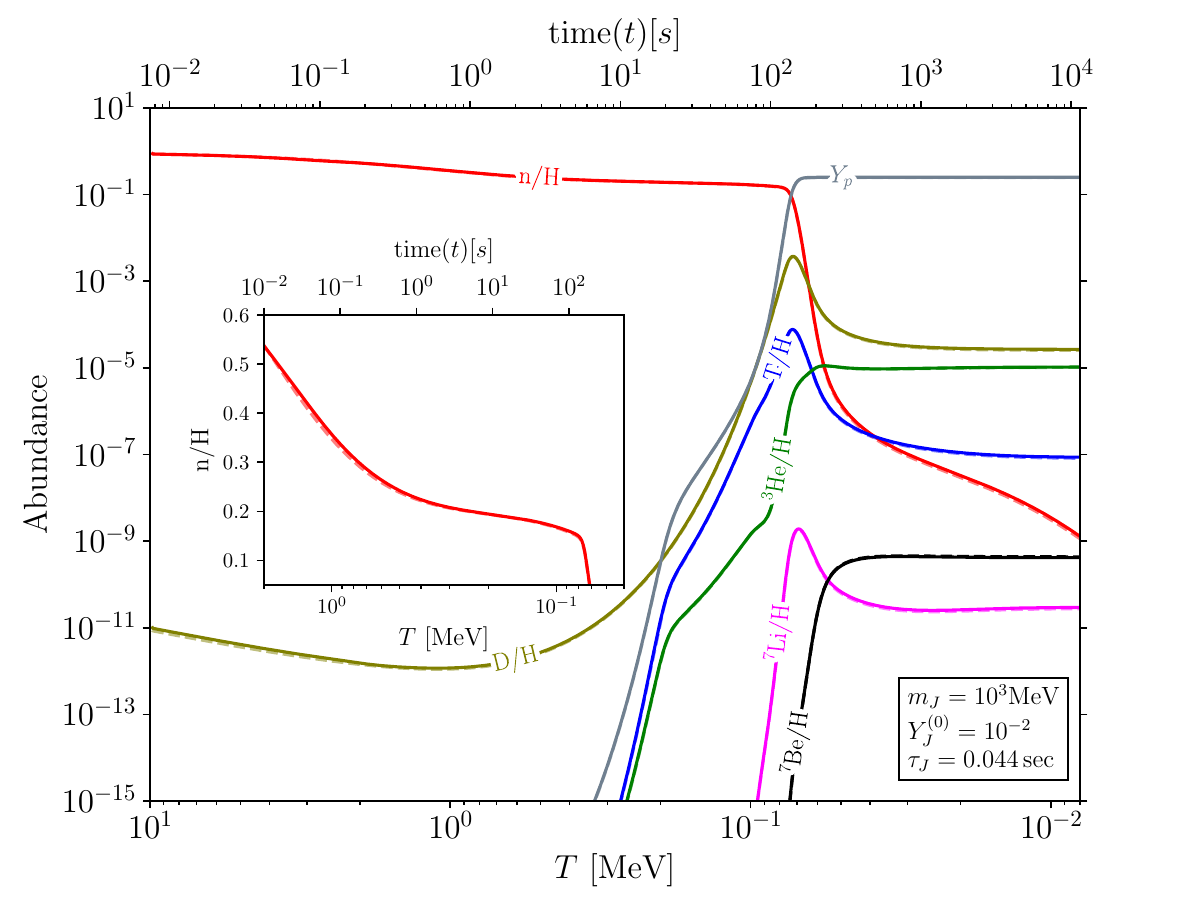}
    \caption{Variation of $n/\H$ (red), $Y_p$ (gray), $\D/\H$ (olive), $\T/\H$ (blue),
    $\ce{^3He}/\H$ (green), $\ce{^7Li}/\H$ (magenta), and $\ce{^7Be}/\H$ (black)
    as a function of temperature $T$ (see the upper tick for the corresponding time).
    In the upper (lower) panel, we take  $m_J=10^3\,\MeV$, $\tau_J = 500\,\sec$, and $Y_J^{(0)} = 10^{-6}$ ($m_J=10^3\,\MeV$, $\tau_J = 0.044\,\sec$, and $Y_J^{(0)} = 10^{-2}$).
    The dashed and solid lines denote the evolution
    for SBBN and SBBN+BSM respectively.}
    \label{Fig:evol_late}
\end{figure}

\section{Results}
\label{sec:results}
\subsection{Evolutions}
In the presence of the majoron decay, the BBN procedure is modified by an interplay of multiple effects as we mentioned previously.
First, additional nuclear reactions are induced by energetic neutrinos, and especially $p\to n$ conversion after the deuterium bottleneck enhances the deuterium abundance as well as all the other elements that can directly be produced from the deuterium.
Second, different heating of $\bnu$ and $e\gamma B$ sectors makes 
$\Tgamma / T_\bnu$
reduced.
As a result, the reaction rates of neutrino induced $n \leftrightarrow p$ conversion
processes are modified (see Appendix.\,\ref{app:n2p_conversion} for their expressions). These modifications result in a shift of the $n/p$ equilibrium
value, and also change its freeze-out temperature in comparison to the SBBN scenario.
Finally, the increased Hubble rate changes the time-to-temperature relation, making all the reactions (including the beta decay) less efficient.

\begin{figure*}[t]
    \centering
   \includegraphics[width = 0.48\textwidth]{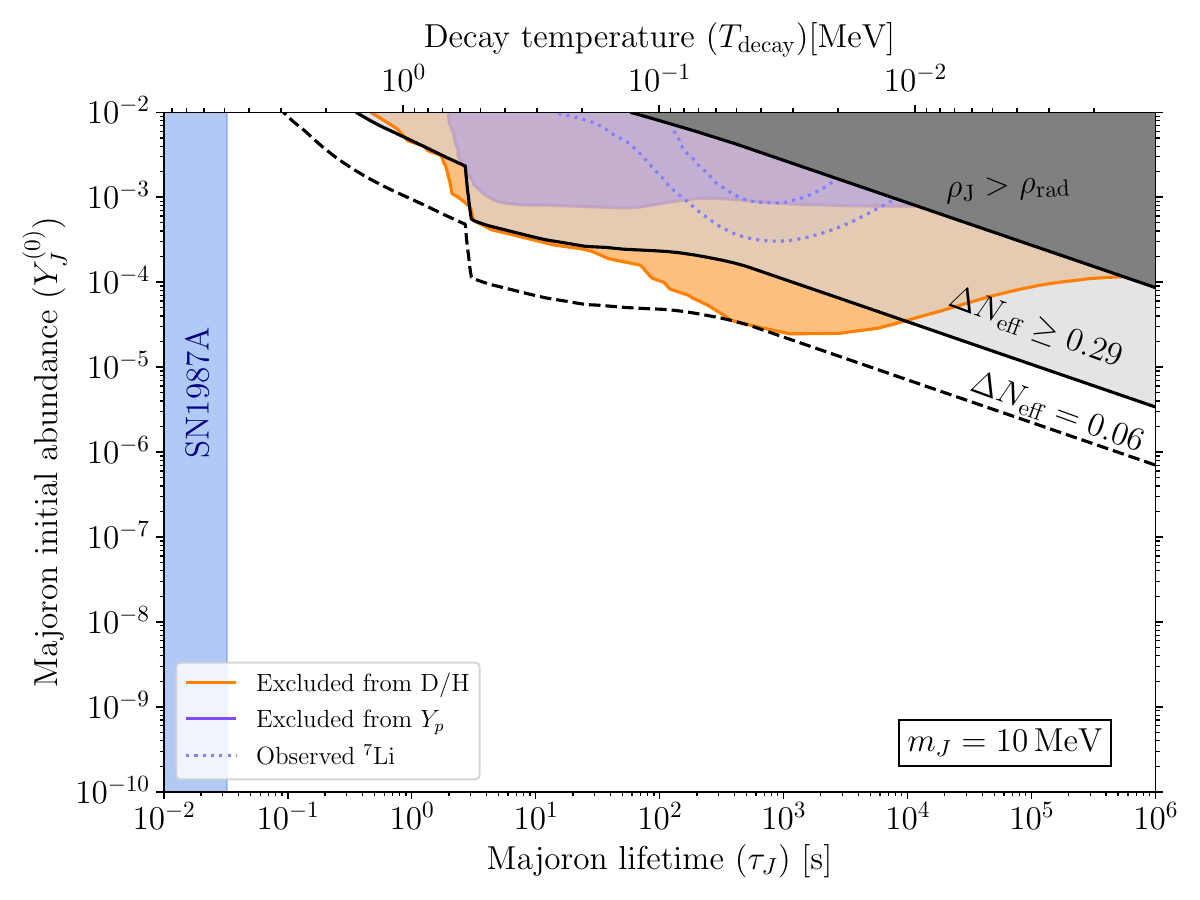}
   \includegraphics[width = 0.48\textwidth]{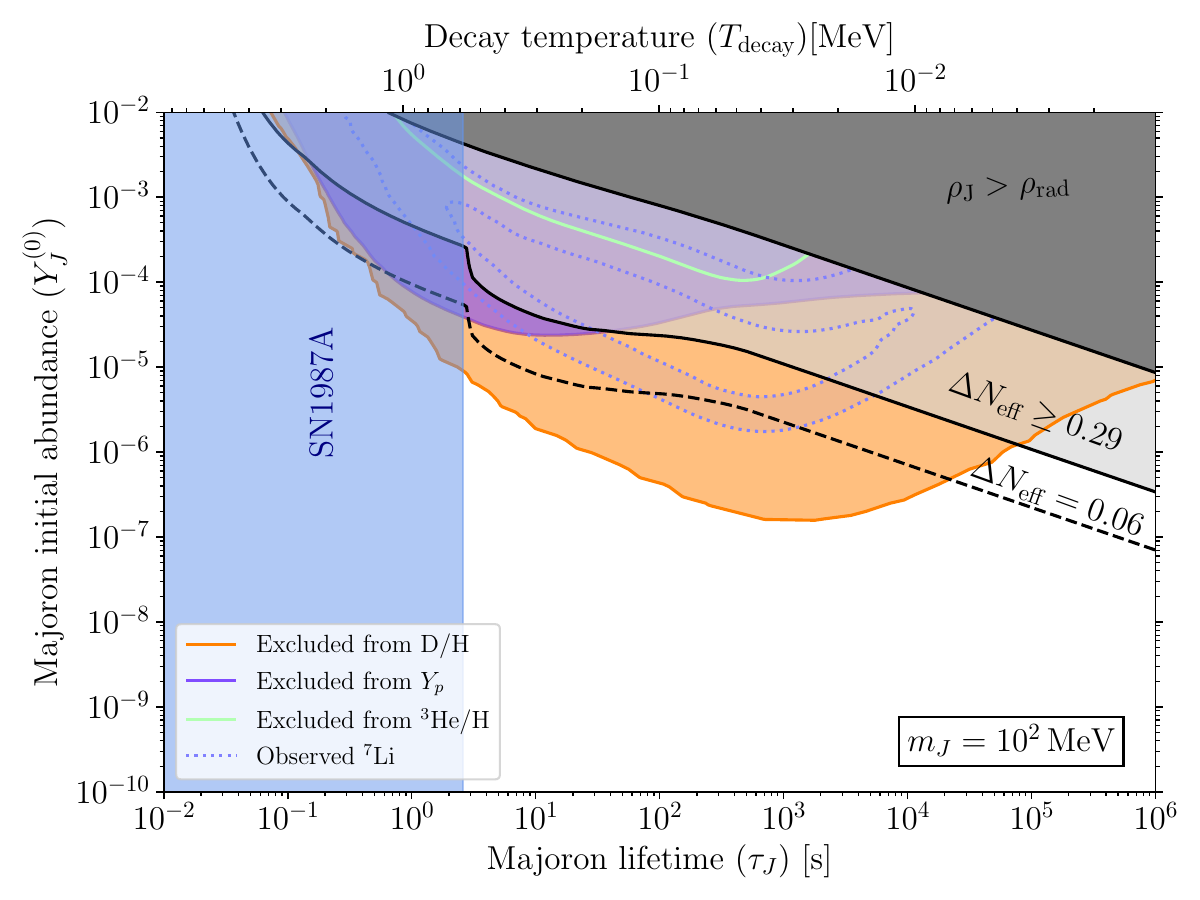}
   \includegraphics[width = 0.48\textwidth]{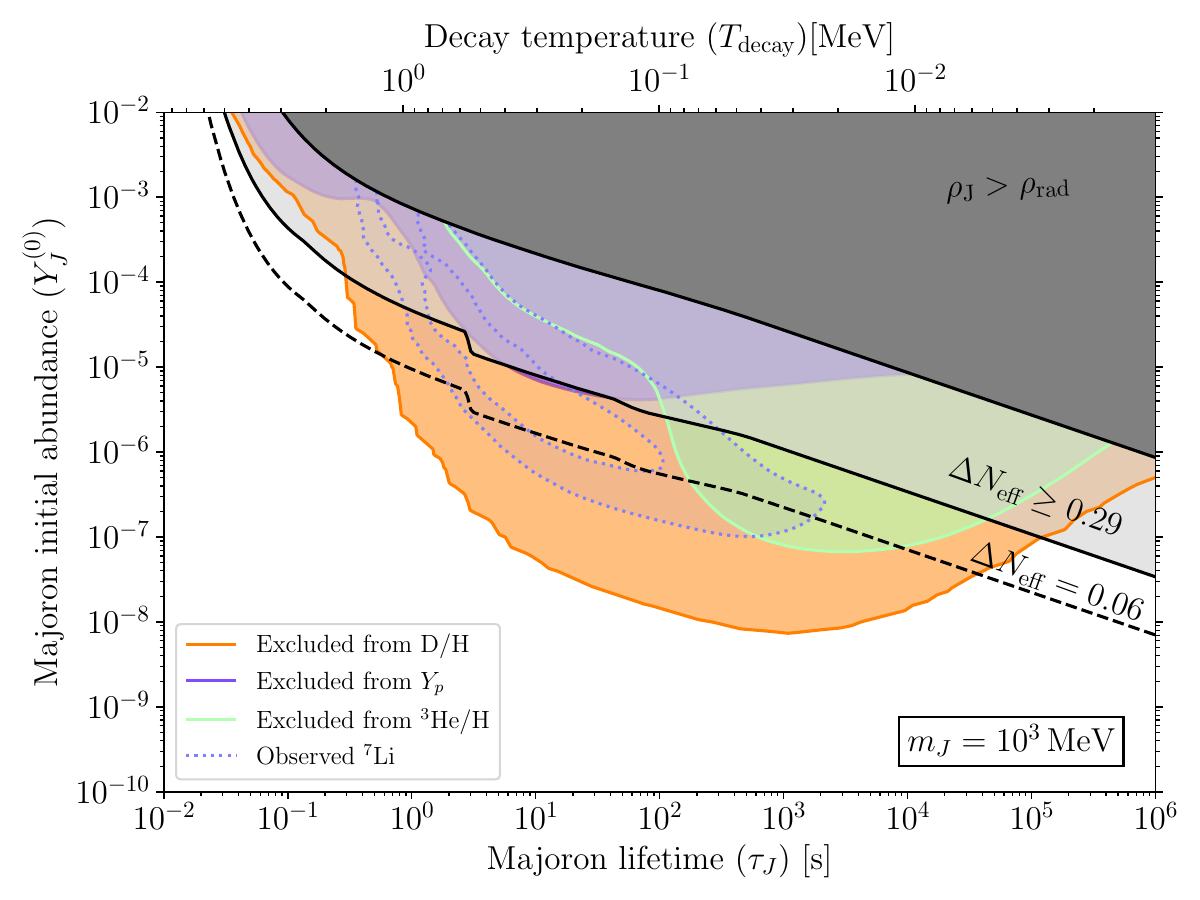}
    \includegraphics[width = 0.48\textwidth]{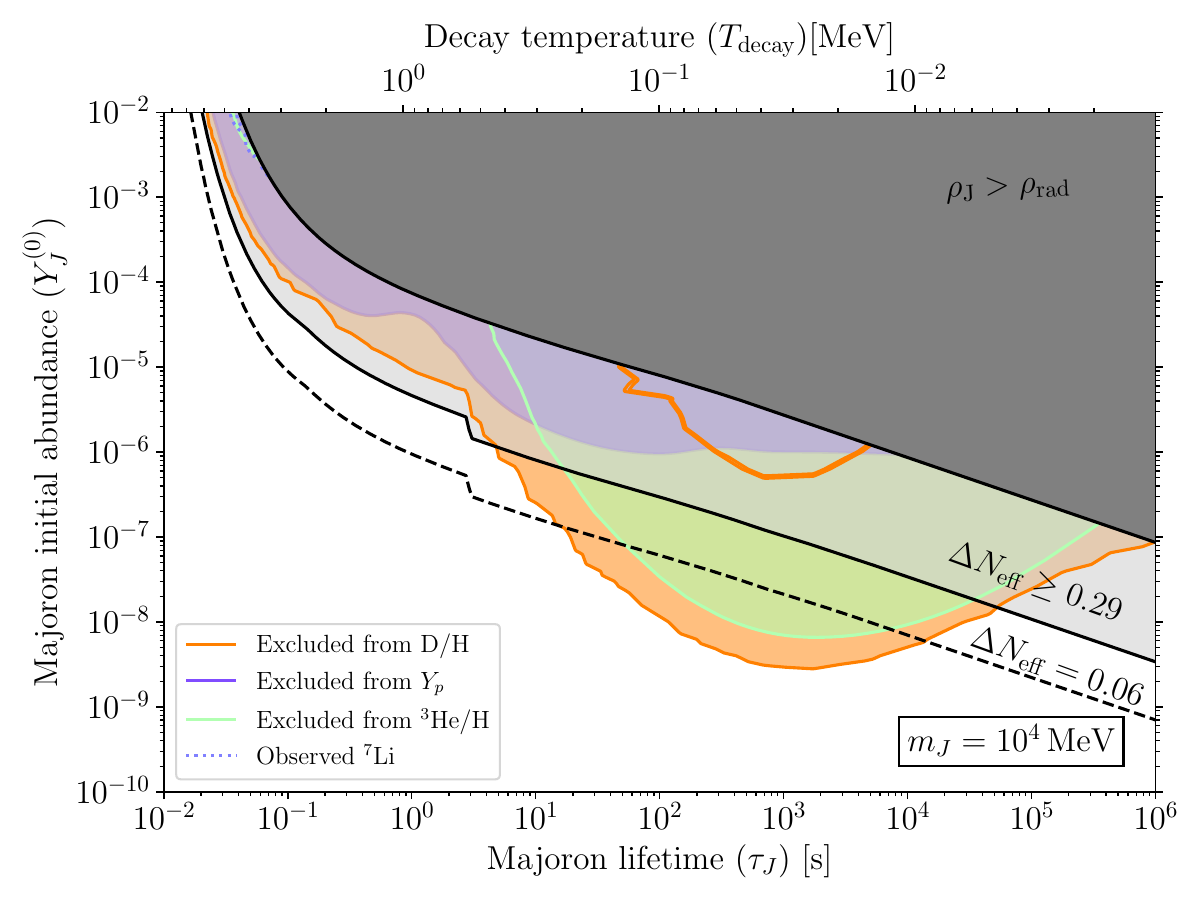}
    \caption{Majoron parameter space in $\tau_J-Y_{J}^{(0)}$ plane for $m_J=10 \,\MeV$ (upper left), $100\,\MeV$ (upper right), 
    $1\,\GeV$ (lower left) and $10\,\GeV$ (lower right).
    Shaded regions are excluded by deuterium (orange), $\ce{^4He}$ (purple), $\ce{^3He}$ (green), $\Delta N_{\rm eff}$ (light gray), and majoron domination (dark gray).
    We show the parameter region (depicted as a blue dotted contour) where the abundance of $\ce{^7Li}$ can be explained, although it is ruled out by other constraints.
    The blue-shaded regions in the upper two panels correspond to the supernova constraint\,\cite{Fiorillo:2022cdq}, which does not exist in the lower two panels because of the heavy majoron mass. 
    }
    \label{Fig:final}
\end{figure*}

The dominant effect is the enhancement of $p\to n$ conversion rate induced by the energetic neutrinos, especially after the deuterium bottleneck $t_\D$ at which the modification of $n\to p$ is negligible because of the small neutron number density compared to the proton number density.
For a large $\tau_J$, most of the energetic neutrinos survive, and 
the abundances of both helium and deuterium are increased as a consequence of additional neutrons.

The other two effects are important when $\tau_J \lsim 1\,\sec$.
The injected neutrinos undergo a large scattering rate expressed by $B_\alpha(\xi,z)$ in Eq.\,\eqref{Eq:Bterm}, which is efficient for a large $T$ and $T_\bnu$.
As $f_\tnu$ is suppressed in this case, the heating effect becomes more important.

To estimate our constraint, we use the values for observed primordial abundances $Y_p=\rho(\ce{^4He})/\rho_b$, $\D/\H$, and $\ce{^7Li}/\H$ recommended in Particle Data Group (PDG)\,\cite{ParticleDataGroup:2022pth}.
We also take the upper bound of $\ce{^3He}/\H$ obtained in the recent analysis presented in Ref.\,\cite{Cooke:2022cvb}:
\begin{table}[H]
\centering
\begin{tabular}{ccc}
\hline
  &  Observation & Ref.
\\ \hline
$Y_p$      &  ~~~$0.245 \pm 0.003$~~~ & \cite{ParticleDataGroup:2022pth}
\\
$\D/\H\times 10^{6}$     & $25.47\pm 0.29$ & \cite{ParticleDataGroup:2022pth}
\\
$\ce{^3He}/\H\times 10^{5}$  &   $<1.09 \pm0.18$ & \cite{Cooke:2022cvb}
\\
$\ce{^7Li}/\H\times 10^{10}$ &    $1.6\pm 0.3$ & \cite{ParticleDataGroup:2022pth}
\\ \hline
\end{tabular}
\end{table}

\noindent
We exclude parameter regions where $Y_p$, $\D/\H$, or $\ce{^3He}/\H$ is out of the $2\sigma$ range.

In our analysis, we do not include the $\ce{^7Li}/\H$ data because it requires a new physics while the majoron cannot solve it as will be shown later; the whole parameter range will be excluded if the $\ce{^7Li}/\H$ data were used.
Likewise, majoron cannot explain the recent measurement of $Y_p$ by the EMPRESS experiment which has a $\sim 1.8\,\sigma$ smaller value compared to the PDG-recommended value\,\cite{Matsumoto:2022tlr}.
We do not use it to avoid an overestimation of our constraint.

In addition, we fix $\eta_B$ by the best-fit value of Ref.\,\cite{Planck:2018vyg}
although including the $\eta_B$ scan can, in principle, make our constraint weaker.
For instance, taking $\eta_B$ to be the upper two-sigma edge of the CMB constraint can reduce the deuterium abundance by a few percent, and therefore the bound can be weaker (the experimental uncertainty is also a few percent).
However, this effect is subdominant compared to other uncertainties such as one-scattering thermalization and instantaneous heating.
Therefore, we do not scan the $\eta_B$ parameter.

In Fig.\,\ref{Fig:evol_late}, we show the evolution of light element abundances for $\tau_J = 500\,\sec$ in the upper panel and  $4.4\times10^{-2}\,\sec$
in the lower panel.
We take the initial abundance $Y_J^{(0)} = 10^{-6}$ (upper) and $10^{-2}$ (lower), while we fix the majoron mass $m_J = 1\,\GeV$.
The dashed lines correspond to the evolutions for the SBBN, i.e. $Y_J^{(0)}=0$, while the solid lines correspond to how they are changed when we include the majoron decay.

For $\tau_J =500\,\sec$ (upper panel), the neutron number density (depicted by the red curve) is increased compared to the SBBN case after the deuterium bottleneck because of the enhanced $p\to n$ conversion. 
It causes the enhancement of the $\D$ abundance (olive) due to the $n+p\to \D +\gamma$ process, and consequently, the abundances of $\D$-sourced elements such as $\T$ (blue), $\ce{^3He}$ (green), and $\ce{^4He}$ (gray) are all enhanced.
On the other hand, the $\ce{^7Be}$ abundance is reduced because of the enhanced $\ce{^7Be}+n \to \ce{^7Li} +p$ reaction.
It accelerates the $\ce{^7Li}+p  \to \ce{^4He} + \ce{^4He}$ process, and
the total $\ce{^7Li} +\ce{^7Be}$ abundance gets reduced, finally.
This effect can be sufficiently strong to fit the observed $\ce{^7Li}$ data, but we find that the parameter space where the $\ce{^7Li}$ problem is 
resolved
is already excluded by the $\D$ constraint.

On the other hand, if neutrinos are injected earlier (as depicted in the bottom panel of Fig.\,\ref{Fig:evol_late}),
the effect of heating and the modified expansion rate is important, which induces sub-processes with different directions.
First, the equilibrium value of the $n/p$ ratio is enhanced as a result of the increased $T_\bnu/T$ ratio.
Second, the freeze-out of the $n/p$ ratio is delayed because the neutrino-induced reactions are enhanced (despite the enhanced Hubble rate).
These two effects give corrections to $Y_p$ with similar size and opposite sign. We find that the final $Y_p$ value gets enhanced, but the impact is small due to the accidental cancellation of these effects.
Finally, the enhanced Hubble rate makes the deuterium bottleneck and the deuterium freeze-out earlier, which enhances $\D/\H$ value.
The bottom panel of Fig.\,\ref{Fig:evol_late} shows an excluded case where $Y_p$ is still within the observed range, but $\D/\H$ is increased too much.

\subsection{Exclusion}
\begin{figure}[t]
    \centering
   \includegraphics[width = 0.43\textwidth]{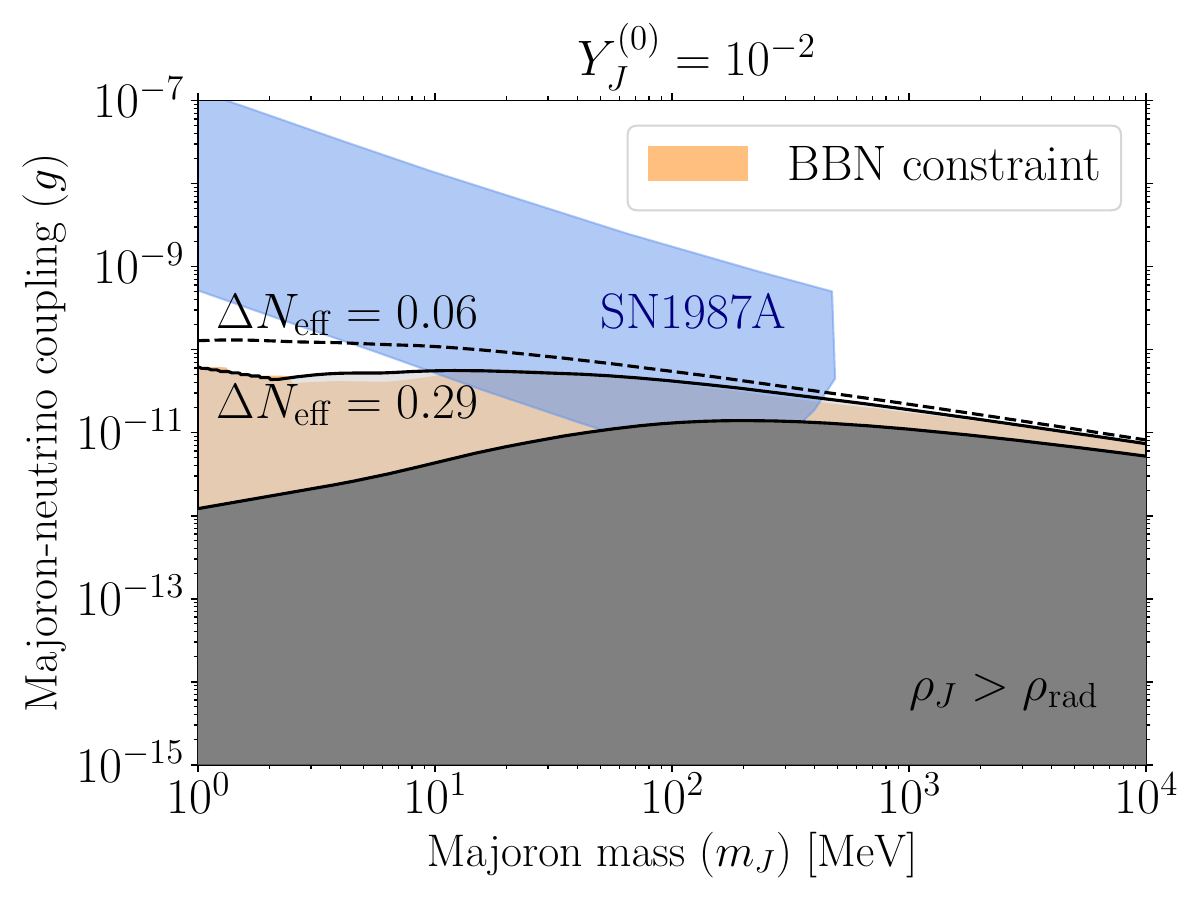}
   \includegraphics[width = 0.43\textwidth]{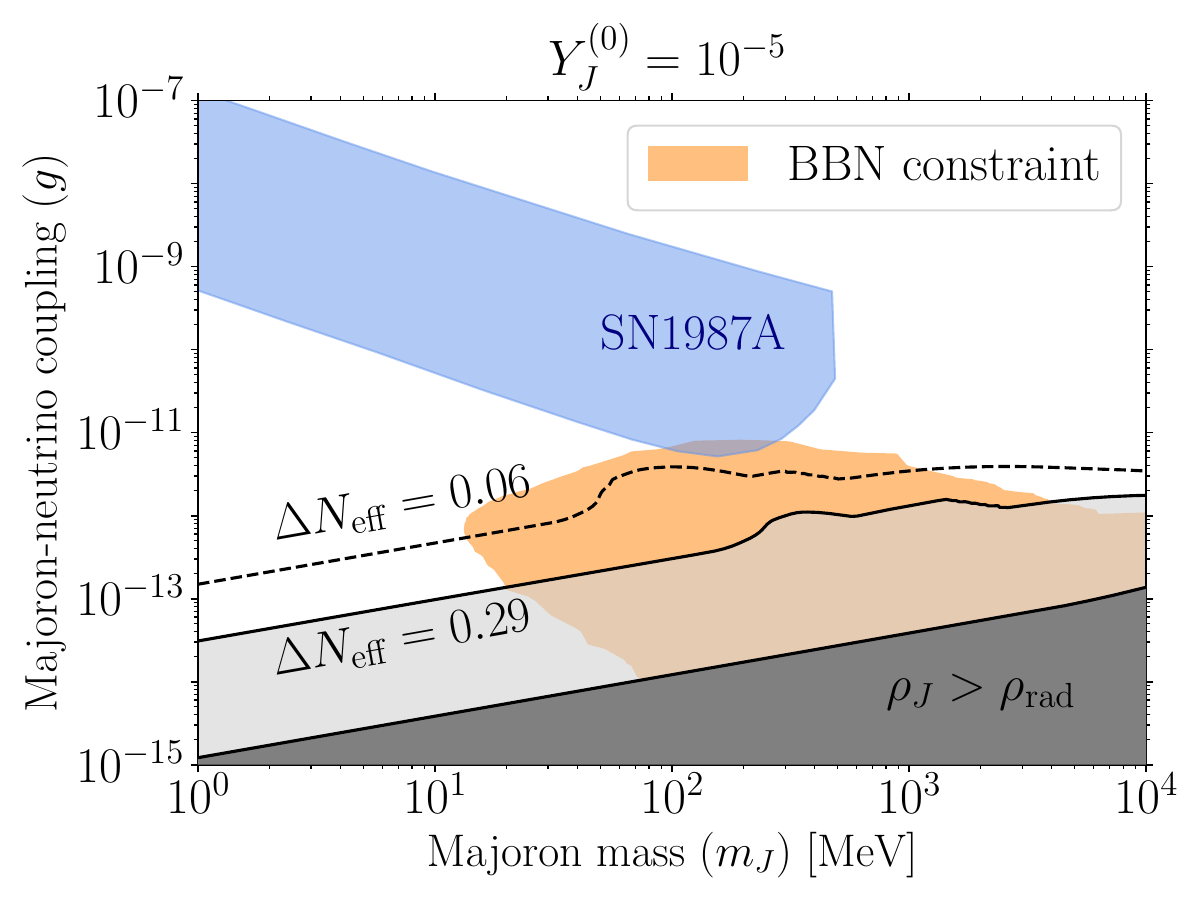}
   \includegraphics[width = 0.43\textwidth]{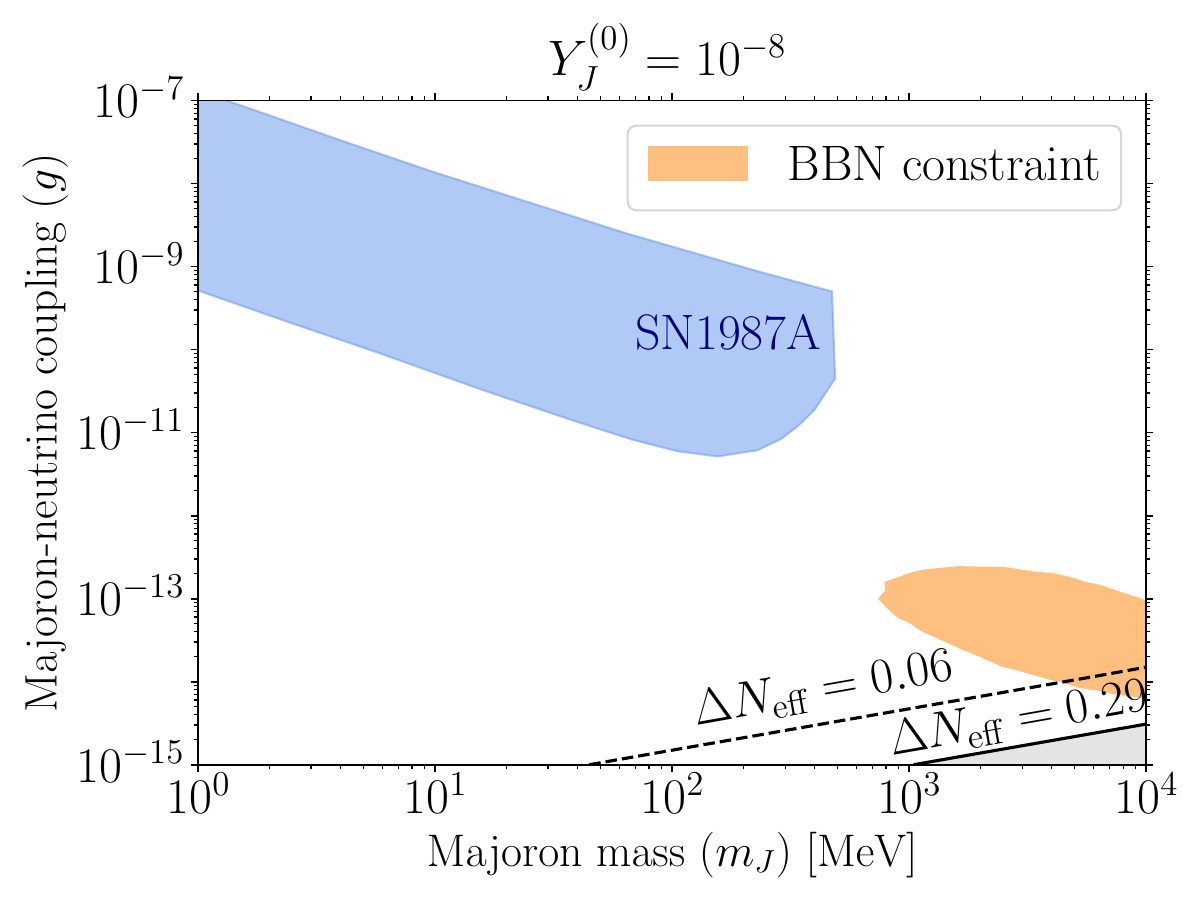}
    \caption{Constraints on Majoron parameter space in $(m_J,\,g)$ plane for $Y_{J}^{(0)}=10^{-2}$ (upper panel), $Y_{J}^{(0)}=10^{-5}$ (middle panel), 
    and $Y_{J}^{(0)}=10^{-8}$ (lower panel).
    In this work, we exclude the shaded regions by the BBN analysis (orange), $\Delta N_{\rm eff}$ (light gray), and the majoron domination (dark gray).
    We also depict the existing supernova constraint (blue)\,\cite{Fiorillo:2022cdq}. 
    }
    \label{Fig:final_mVSg}
\end{figure}

Our constraints are summarized in Fig.\,\ref{Fig:final} and \ref{Fig:final_mVSg}.
Fig.\,\ref{Fig:final} is in the parameter space of $\tau_J$ and $Y_J^{(0)}$ for $m_J=10\,\MeV$, $100\,\MeV$, $1\,\GeV$ and $10\,\GeV$, while Fig.\,\ref{Fig:final_mVSg} is their projection to the $m_J$ and $g$ space 
for $Y_J^{(0)}=10^{-2},$ $10^{-5}$ and $10^{-8}$.
In Fig.\,\ref{Fig:final}, the orange regions depict the strong constraint from the $\D$ abundance, while the green and purple contours correspond to $\ce{^3He}$ and $\ce{^4He}$ bounds, respectively (although they are weaker than the $\D$ constraint).
We also show the $\Delta N_{\rm eff}$ constraint\footnote{
We take the current limit on $N_{\rm eff}$ as $2.99^{+0.34}_{-0.33}$ at the $95\%$ confidence level\,\cite{Planck:2018vyg} while we take the SM value of $N_{\rm eff}$ by $3.04$\,\cite{Mangano:2001iu,Mangano:2005cc,deSalas:2016ztq, Bennett:2019ewm, Cielo:2023bqp}. Therefore, the upper bound corresponds to $\Delta N_{\rm eff}<0.29$.
}
from the CMB analysis\,\cite{Planck:2018vyg} 
by the light gray and the future sensitivity of CMB Stage-4\,\cite{CMB-S4:2016ple} 
by the dashed line. The $\Delta N_{\rm eff}$ constraint becomes stronger than the $\D$ constraint for a short lifetime.
Note that the wiggles/kinks represent the uncertainty of our estimation which comes from various step functions in our analysis.
We also show the SN1987A constraint of Ref.\,\cite{Fiorillo:2022cdq} in the figures (see also Refs.\,\cite{Kolb:1987qy,Choi:1987sd,Choi:1989hi,Farzan:2002wx,Shalgar:2019rqe,Akita:2022etk,Akita:2023iwq,Fiorillo:2023cas,Fiorillo:2023ytr}).
 
Our framework breaks down when the majoron energy density dominates (shaded by the dark gray in Fig.\,\ref{Fig:final} and \ref{Fig:final_mVSg}).
If this happens, the reheating temperature after majorons' decay can be approximated to the decay temperature of the majoron, and therefore, $T_{\rm decay} \lsim \MeV$ is strongly ruled out.
However, obtaining a precise lower bound of reheating temperature matters for $m_J>\GeV$, as it can happen with $T_{\rm decay} \gsim \MeV$ (see Fig.\,\ref{Fig:final}). 
Although it requires a more careful and sophisticated estimation of neutrino distribution, 
we expect the result will be stronger than the cases of radiative or hadronic channel\,\cite{Kawasaki:1999na,Kawasaki:2000en,Hannestad:2004px,Ichikawa:2005vw,deSalas:2015glj,Hasegawa:2019jsa} because thermalization of the plasma starting from neutrinos should be much less efficient.

Since the initial abundance of majoron $Y_J^{(0)}$ is sensitive to the history of the universe, we take a wide range of $Y_J^{(0)}=10^{-2},\,10^{-5}$ and $10^{-8}$, and show the constraints in $(m_J,\,g)$ plane in 
Fig.\,\ref{Fig:final_mVSg}.
$Y_J^{(0)} = 0.28/g_{*s}(T_{\rm FO}) \simeq 10^{-2}$ represents the case where the majorons are maximally produced and frozen-out at $T_{\rm FO} \gg m_J$.
Such a case can easily be realized when the universe undergoes the $B-L$ phase transition.
As shown in the top panel in Fig.\,\ref{Fig:final_mVSg}, the BBN and $\Delta N_{\rm eff}$ constraints are comparable to each other, and the constraint from the reheating temperature of majoron dominated era excludes the bottom region of the parameter space.

On the other hand, if the reheating temperature after the inflation is much less than the $B-L$ symmetry breaking scale $f_J$, it is extremely difficult for majorons to be fully thermalized due to the intrinsically small coupling, $g_{\alpha \beta}\simeq m_{\nu, \alpha \beta}/f_J$, and $Y_J^{(0)}$ can be arbitrarily small depending on the UV models (see, e.g. Ref.\,\cite{Frigerio:2011in, Garcia-Cely:2017oco, Brune:2018sab, Abe:2020dut, Manna:2022gwn, Li:2023kuz}).  
In the middle and bottom panel of Fig.\,\ref{Fig:final_mVSg}, we take $Y_J^{(0)}= 10^{-5}$ and $10^{-8}$ as references of nonthermal scenarios.

\section{Conclusion}
\label{sec:conclu}
In this paper, we have estimated the BBN constraint on majoron in the mass range $\MeV \le m_J \le 10\,\GeV$.
When $\tau_J \gsim 1\,\sec$, the decay of majorons leaves energetic neutrinos, and they contribute to an additional $p\to n$ conversion.
On the other hand, the effects of heating and the modified Hubble rate result in a relatively mild constraint at $\tau_J\lsim 1\,\sec$.
We find that, in both cases, the deuterium abundance provides the strongest constraint among the measured primordial light elements.

The additional neutrons due to the injected neutrinos reduce the $\ce{^7Be}$ abundance (and thus $\ce{^7Li}$ at present).
However, the parameter region 
that explains the present observation on the
primordial $\ce{^7Li}$ abundance
is ruled out by the strong constraint from the deuterium abundance.

We also estimate other cosmological constraints such as the $\Delta N_{\rm eff}$ bound from the CMB analysis and the reheating temperature bound on majoron dominated scenario.
For the maximally thermalized scenario with 
$Y_J^{(0)}\simeq 10^{-2}$,
the BBN constraint is comparable to the $\Delta N_{\rm eff}$ bound.
On the other hand, our BBN analysis rules out a distinctive region of parameter space for nonthermal majoron scenarios with $Y_J^{(0)} \ll 10^{-2}$.

Exploring the higher mass region requires more careful consideration.
First of all, one should include processes of neutrino annihilation into heavier particles such as $\nu \bar \nu \to \mu^+ \mu^-,\,  \pi^+\pi^-, \cdots$.
These channels easily mess up the neutron-to-proton ratio, and thus we expect a stronger constraint will be put on the short lifetime.
Moreover, heavy majorons can directly decay to SM fermions via one loop level\,\cite{Heeck:2019guh}, where the branching ratio is roughly $10^{-4} m_\nu^2 m_f^2/g^2 v_h^4$ for the Higgs vacuum expectation value $v_h=246\ \GeV$, effecive neutrino mass $m_\nu\sim 0.1\,{\rm eV}$, and the fermion mass $m_f$.
These additional decay channels would be more dangerous than the neutrino mode although the branching ratio is small.

Our analysis can be further improved by a more realistic treatment of scattered neutrinos.
This is crucial, especially for $\tau_J \lsim 1\,\sec$ where the scattering term \eqref{Eq:Bterm} is efficient.
However, since it takes a significantly large amount of computational resources, we leave it for future work.

\vskip 1em
\noindent
{\bf Acknowledgement:}
This work was supported by IBS under the project code IBS-R018-D1 and IBS-R031-D1.
The work of SG and CSS was supported by the National Research Foundation of Korea (NRF-2022R1C1C1011840). CSS also acknowledges support from (NRF-2022R1A4A5030362).
SG would like to acknowledge the hospitality
of the Institute for Basic Science during the course of this work. SG thanks Ofelia Pisanti
for a few email conversations related to the \texttt{PArthENOPE} code.

\begin{appendix}

\begin{widetext}

\section{The scattering term of injected neutrinos}
\label{App:scattering}
In the estimation of ${\cal C}_{\tnu a \to bc}$ 
in Eq.\,\eqref{Eq:C_scattering}, we approximate all the external particles are massless so that we can simply factor out the energy dependence of the corresponding cross section as
\bal
\sigma_{\nu_{\rm nt, \alpha} a \to bc} = \zeta_{abc} 
\frac{ G_F^2 E_{\rm cm}^2 }{\pi},
\eal
where $E_{\rm cm}$ is the center of mass energy and $\zeta_{abc}$ is a coefficient as we summarize in Table.\,\ref{Tab:zeta_e} which agrees with Ref.\,\cite{Hannestad:1995rs}.
Taking zero neutrino masses is, of course, valid since $T\gg m_\nu$.
Taking $m_e=0$ at $T \gg m_e$ is a good approximation, but the uncertainty becomes order one when $T \simeq m_e$.
At $T < m_e$, the interaction rates involving $e^+$ or $e^-$ are suppressed by the Boltzmann factor, so we turn off the corresponding collision term by using the step function.

\begin{table}[t]
\begin{center}
\begin{tabular}{|c|c|}
 \hline
process ($\tnu a \to bc$) & ~~$\zeta_{abc}$~~
\\ \hline
$\nu_e + \bar \nu_e \to \nu_e + \bar \nu_e $ 
& 2/3 \\  
$\nu_e + \nu_e \to \nu_e + \nu_e$
& 1  \\  
$\nu_e + \nu_i \to \nu_e + \nu_i$ 
& 1/2 \\ 
$\nu_e + \bar \nu_i \to \nu_e + \bar \nu_i$ 
& 1/6 \\ 
$\nu_e + \bar \nu_e \to \nu_i + \bar \nu_i$
& 1/6 \\ \hline \hline
$\nu_e + e^- \to \nu_e + e^-$
& $(C_A^2+C_A C_V+C_V^2)/3$ \\ 
$\nu_e + e^+ \to \nu_e + e^+$
& $(C_A^2-C_A C_V+C_V^2)/3$ \\ 
$\nu_e + \bar \nu_e \to e^- + e^+$
& $(C_A^2+C_V^2)/3$ \\ 
\hline \hline
$\nu_i + e^- \to \nu_i + e^-$ 
& $[3(C_A+C_V-2)^2+(C_A-C_V)^2]/12$  \\ 
$\nu_i + e^+ \to \nu_i + e^+$ 
& $[(C_A+C_V-2)^2+3(C_A-C_V)^2]/12$ \\ 
$\nu_i + \bar \nu_i \to e^- + e^+$ 
& $[(C_A+C_V-2)^2+(C_A-C_V)^2]/6$ \\ \hline
\end{tabular}
\end{center}
\caption{$\zeta_{abc}$ for $\nu_e$, where we take $m_e=0$.
$C_V=\frac{1}{2}+2\sin^2 \theta_W$ and $C_A=\frac{1}{2}$.}
\label{Tab:zeta_e}
\end{table}

With taking the M{\o}ller velocity\,\cite{Gondolo:1990dk} $v=((p_1\cdot p_2)^2 - m_1^2 m_2^2)^{1/2}/(E_1 E_2)=1-\cos \theta$, $E_{\rm cm}^2 = 2 E_\tnu E_a (1-\cos \theta)$, and $f_a=1/(e^{E_a/T_a}+1)$, we obtain 
\bal
{\cal C}_{\tanu a \to bc}
&=
-2 
f_\tanu
\int d\Pi_a   \,   ( \sigma_{\tanu a\to bc} \,v) \, E_a \, f_a
= 
g_a \, 
\zeta_{abc} \, G_F^2 E_\tnu
\frac{4f_\tanu}{3\pi^3}
\int dE_a 
\,
E_a^3 f_a(E_a)
\nonumber \\
&= 
\frac{7\pi}{90}
\, G_F^2 E_\tnu T^4_a 
\, f_\tanu
g_a \,
\zeta_{abc} 
\eal
where $g_a$ is the spin-degeneracy,
$g_\nu=1$ and $g_e=2$ and $\alpha=e,\mu,\tau$.
Note that the symmetry factor $1/(1+\delta_{a\tanu})$ is canceled by the coefficient of $2 f_{\bnu} f_{\tanu}$ that comes from $f_\nu f_\nu = f_\bnu^2 + 2 f_\bnu f_\tanu + f_\tanu^2$.
Then, $B_\alpha(\xi,z)$ in Eq.\,\eqref{Eq:Boltzmannn_AB} is given by
\bal
B_\alpha(\xi, z) =
\frac{7\pi G_F^2 m_e^5}{90H}\frac{\xi}{z^6}
\Bigg[&
\zeta_{\alpha 1} \, \theta(T-m_e)
+ 
\left(\frac{T_\bnu}{T}\right)^4
\Big[\zeta_{\alpha 2}+\zeta_{\alpha 3} \, \theta(E_\tnu T_\bnu -m_e^2)
\Bigg]
\Bigg]\,,
\eal
where 
\bal 
&\zeta_{e1}=\frac{4}{3}(C_A^2+C_V^2),\ \zeta_{e2} =\frac{10}{3},\ 
\zeta_{e3}=\frac{1}{3}(C_A^2+C_V^2), \  
\zeta_{\mu 1} = \zeta_{\tau 1}= \frac{1}{3} \Big[(C_A+C_V-2)^2+(C_A-C_V)^2\Big], 
\nn\\
&\zeta_{\mu 2} = \zeta_{\tau 2} = \zeta_{e 2}, \ 
\zeta_{\mu 3}=\zeta_{\tau 3}=\frac{1}{3} \Big[(C_A+C_V-2)^2+(C_A-C_V)^2\Big].
\eal

The interactions in the $B$ term are directly related to  $\Gamma(\tanu \to \nu, e)$ of Eq.\,\eqref{Eq:W1} and \eqref{Eq:W2}. 
The analytical expressions of $\Gamma(\tanu \to \nu, e)$  for different flavors of $\tnu$ are given by
\bal
\Gamma(\nu_{{\rm nt}, e} \to \nu) &=
\langle \,
	\sigma v (\tnu \nu \to \nu \nu) 
	+\sigma v (\tnu \bar \nu \to \nu \bar \nu) 
\, \rangle n_\nu 
+\frac{1}{2}\langle 
	\sigma v (\tnu e^\pm \to \nu e^\pm) 
\rangle n_e 
\nn
\\
&\simeq \frac{7\pi}{90}
G_F^2 E_\tnu T_{\bnu,*}^4
\left[ 
    \frac{10}{3}+
  \frac{2}{3} 
    \left( \frac{\Tgammastar}{T_{\bnu,*}}\right)^4
    (C_A^2+C_V^2)\theta(\Tgammastar-m_e) 
\right],
\\
\Gamma(\nu_{{\rm nt}, \mu} \to \nu)&=\Gamma(\nu_{{\rm nt}, \tau} \to \nu)=\langle \,
	\sigma v (\tnu \nu \to \nu \nu) 
	+\sigma v (\tnu \bar \nu \to \nu \bar \nu) 
\, \rangle n_\nu 
+\frac{1}{2}\langle 
	\sigma v (\tnu e^\pm \to \nu e^\pm) 
\rangle n_e 
\nn
\\
&\simeq \frac{7\pi}{90}
G_F^2 E_\tnu T_{\bnu,*}^4
 \left[ 
    \frac{10}{3}+
    \frac{1}{3} 
    \left( \frac{\Tgammastar}{T_{\bnu,*}}\right)^4
    [(C_A-C_V)^2+(C_A+C_V-2)^2]\theta(\Tgammastar-m_e) 
\right],
\\
\Gamma(\nu_{{\rm nt}, e} \to e)&=
\frac{1}{2}\langle 
	\sigma v (\tnu e^\pm \to \nu e^\pm) 
\rangle n_e 
+
\langle 
	\sigma v (\tnu \bar \nu \to e^+ e^-)
\rangle n_\nu
\nn
\\
&\simeq \frac{7\pi}{90} G_F^2 E_\tnu \Tgammastar^4   (C_A^2+C_V^2)  \left[
  \frac{2}{3} \theta(\Tgammastar-m_e) 
    + \frac{1}{3} \left(\frac{T_{\bnu,*}}{\Tgammastar}\right)^4\theta(E_\tnu T_{\bnu,*} -m_e^2)
\right],
\\
\Gamma(\nu_{{\rm nt}, \mu} \to e )&=\Gamma(\nu_{{\rm nt}, \tau} \to e)=
\frac{1}{2}\langle 
	\sigma v (\tnu e^\pm \to \nu e^\pm) 
\rangle n_e 
+
\langle 
	\sigma v (\tnu \bar \nu \to e^+ e^-)
\rangle n_\nu
\nn
\\
&\simeq 
\frac{7\pi}{90} G_F^2 E_\tnu \Tgammastar^4  \Big[(C_A-C_V)^2+(C_A+C_V-2)^2\Big]
 \left[
  \frac{1}{3} \theta(\Tgammastar-m_e) 
    + \frac{1}{6} \left(\frac{T_{\bnu,*}}{\Tgammastar}\right)^4\theta(E_\tnu T_{\bnu,*} -m_e^2)
\right].
\eal

\section{Cross sections of $\tnu$ involving nuclear reactions}
\label{App:cross_section}
The scattering cross section of $\tnu$ with $n$  and $p$ for $E_{\tnu} < 300 \,{\MeV}$ is given by \cite{Strumia:2003zx}
\bal
&\sigma_{\tnu n \to p e^-} \simeq 9.52 \times 10^{-44} {\rm cm}^2 \dfrac{E_e}{\MeV} \dfrac{p_e}{\MeV}  \label{Eq:n2p_lowE},\\
&\sigma_{\tnubar p \to n e^+} \simeq 10^{-43} {\rm cm^2} \dfrac{E_e}{\MeV} \dfrac{p_e}{\MeV} 
\left(\dfrac{E_{\tnu}}{\MeV}\right)^\gamma ,\label {Eq:p2n_lowE} 
\eal
where
\bal
\gamma = -0.07056 + 0.02018 \ln\left(\dfrac{E_{\tnu}}{\MeV}\right) - 0.001953 \ln\left(\dfrac{E_{\tnu}}{\MeV}
\right)^3\!.
\eal

In  \eqref{Eq:n2p_lowE}, $E_e = E_{\tnu} + m_n - m_p$ whereas $E_e = E_{\tnu} - (m_n - m_p)$ in \eqref{Eq:p2n_lowE} and 
$p_e = \sqrt{E_e^2 - m_e^2}$.
For $E_{\tnu}\ge 300 \, {\MeV}$, the scattering cross sections of $\tnu$ with $n$ and $p$ are given in Table \ref{Tab:nucleon_cross}.

In our analysis, we have considered the interactions of $\tnu$ with deuterium $(\rm D)$ and helium $({}^4\rm He)$ and the relevant cross-sections are tabulated 
in Table \ref{Tab:deuterium_cross} and Table \ref{Tab:helium_cross} respectively. The full tables can be found in \cite{Tatara:1990eb} and \cite{Yoshida:2008zb}.

For highly energetic nonthermal neutrinos, the data is not available and in this case, we have extrapolated the scattering cross-section of nonthermal neutrinos with $\D$ and $\ce{^4He}$.
The extrapolation has been performed using the following formula.
\bal
\sigma = \sigma (E_0) \left(\dfrac{E_{\tnu}}{E_0}\right)^2 \left[\dfrac{E_0^2 + \Lambda^2}{E_{\tnu}^2 + \Lambda^2}\right]\,\,,
\eal
where $E_0$ is the maximum value of the nonthermal neutrino energy up to which the data is available and $\sigma (E_0)$ is the cross section at $E_0$. Here we have considered
$\Lambda = 1\, \GeV$.

\begin{table}[t] 
\begin{center}
\begin{tabular}{|c | c c|}
    \hline
     $E_{\tnu}$[MeV]&  ${\tnubar p \to n e^+} $&  ${\tnu n \to p e^-} $   \\
     \hline
    300   &  1.48 &  5.37  \\ 
    350   &  1.71 &  6.36  \\ 
    400   &  1.93 &  7.22  \\ 
    450   &  2.15 &  7.94  \\ 
    500   &  2.36 &  8.53 \\ 
    550   &  2.57 &  9.02  \\ 
    600   &  2.77 &  9.42  \\ 
    650   &  2.97 &  9.73  \\ 
    700   &  3.16 &  9.99  \\ 
    750   &  3.34 &  10.19  \\ 
    800   &  3.51 &  10.35  \\ 
    850   &  3.67 &  10.47  \\
    900   &  3.83 &  10.57  \\ 
    950   &  3.98 &  10.64  \\ 
    1000   &  4.12 &  10.69  \\ \hline
\end{tabular}
\caption{Scattering cross-sections 
of nonthermal neutrinos with nucleons in units of femtobarn (fb).}
\label{Tab:nucleon_cross}
\end{center}
\end{table}

\begin{table}[t]
    \begin{center}
    \begin{tabular}{|c | c c c c|}
    \hline
    $E_{\tnu}$[MeV] & $\D (\tnu, \nu) n p$ &  $\D ({\tnubar}, {\bar \nu}) n p$ & $\D (\tnu, e^-) p p$ & $\D ({\tnubar},e^+) n n$ \\ \hline
    4 & $3.07 \times 10^{-5}$  & $3.02 \times 10^{-5}$  & $1.58 \times 10^{-4}$ & $0.00$  \\ 
  10  & $1.10 \times 10^{-3} $  & $1.05 \times 10^{-3}$  & $2.71 \times 10^{-3}$ & $1.23 \times 10^{-3}$ \\ 
  50  & 
  $5.91 \times 10^{-2}$  
  & $4.52 \times 10^{-2}$  & $0.134$                 & $7.29 \times 10^{-2}$  \\ 
 100  & $0.262$                 & $0.158$                  & $0.635$                 & $0.239$   \\
 170  & $0.706$                & $0.330$                  & $1.82$                 & $0.425$    \\ \hline
    \end{tabular}
    \caption{Scattering cross sections of nonthermal neutrinos with 
    deuterium in units of femtobarn (fb).}
    \label{Tab:deuterium_cross}
    \end{center}
    \end{table}

\begin{table*}[t] 
    \begin{center}
    \begin{tabular}{|c| c c c c c|}
    \hline
    $E_{\tnu}$[MeV] & $\ce{^4He}(\tnu, \nu)p {}^3\H$ &  $\ce{^4He} (\tnu, \nu) n {}\ce{^3He}$ & $\ce{^4He} (\tnu, \nu) \D \D$ & $\ce{^4He} (\tnu, e^-) p {}\ce{^3He} $ & $\ce{^4He} ({\tnubar}, e^+) n {}^3\H$ \\ \hline
    50 & $1.80 \times 10^{-3}$ & $1.74 \times 10^{-3}$    & $7.22 \times 10^{-5}$ & $8.96 \times 10^{-3}$ & $5.99 \times 10^{-3}$ \\ 
    75 & $1.40 \times 10^{-2}$ & $1.36\times 10^{-2}$ & $1.12\times 10^{-3}$ & $8.31\times 10^{-2}$ & $4.18\times 10^{-2}$ \\ 
    100 & $4.76\times 10^{-2}$ & $4.63\times 10^{-2}$ & $3.57\times 10^{-3}$ & $3.26\times 10^{-1}$ & $1.26\times 10^{-1}$ \\ 
    150 & $1.89\times 10^{-1}$ & $1.85\times 10^{-1}$ & $1.52\times 10^{-2}$ & $1.65$ & $4.10\times 10^{-1}$  \\ 
    180 & $2.98\times 10^{-1}$ & $2.92\times 10^{-1}$ & $2.77\times 10^{-2}$ & $2.95$ & $6.02\times 10^{-1}$ \\ \hline
    \end{tabular}
    \caption{Scattering cross sections of nonthermal neutrinos with 
    $\ce{^4He}$ in units of femtobarn (fb).}
    \label{Tab:helium_cross}
    \end{center}
\end{table*}

\section{Modified $n \leftrightarrow p$ conversion rate due to neutrino heating}
\label{app:n2p_conversion}
The quantity $x_{np}$ and $x_{pn}$ is defined as
\bal
x_{np} = \dfrac{{\Gamma_{n \nu_e \to p e^-} (T^\prime_\bnu) + \Gamma_{n e^+ \to p \bar{\nu}_e} (T)}}
{{\Gamma_{n \nu_e \to p e^-} (T_\bnu) + \Gamma_{n e^+ \to p \bar{\nu}_e} (T)}}
\,\,,\nn\\
x_{pn} = \dfrac{{\Gamma_{p \bar{\nu}_e \to n e^+} (T^\prime_\bnu) + \Gamma_{p e^- \to n \nu_e} (T)}}
{{\Gamma_{p \bar{\nu}_e \to n e^+} (T_\bnu) + \Gamma_{p e^- \to n \nu_e} (T)}}\,\,,
\eal
where $T^\prime_\bnu = T_\bnu (1 + \Delta T_\bnu/T_\bnu)$. The explicit forms of the reaction rates 
(neglecting the Pauli blocking factor for the final state fermion)
are given by
\bal
&\Gamma_{n \nu_e \to p e^-} (T_\bnu) \simeq \dfrac{1 + 3 g_A^2}{2 \pi^3} G_F^2 Q^5 
\int_1^\infty dq\sqrt{1 - \dfrac{(m_e/Q)^2}{q^2}}\dfrac{\, q^2 (q-1)^2}{1 + e^\frac{Q (q-1)}{T_\bnu}}\,,\\
&\Gamma_{n e^+ \to p \bar{\nu}_e} (T) \simeq \dfrac{1 + 3 g_A^2}{2 \pi^3} G_F^2 Q^5 
\int_{-\infty}^{-m_e/Q} dq\sqrt{1 - \dfrac{(m_e/Q)^2}{q^2}}\dfrac{\, q^2 (q-1)^2}{1 + e^\frac{-Q q}{T}}\,,\\
&\Gamma_{p \bar{\nu}_e \to n e^+} (T_\bnu) \simeq \dfrac{1 + 3 g_A^2}{2 \pi^3} G_F^2 Q^5 
\int_{-\infty}^{-m_e/Q} dq\sqrt{1 - \dfrac{(m_e/Q)^2}{q^2}}\dfrac{\, q^2 (q-1)^2}{1 + e^\frac{-Q (q-1)}{T_\bnu}}\,,\\
&\Gamma_{p e^-\to n \nu_e} (T) \simeq \dfrac{1 + 3 g_A^2}{2 \pi^3} G_F^2 Q^5 
\int_1^\infty dq\sqrt{1 - \dfrac{(m_e/Q)^2}{q^2}}\dfrac{\, q^2 (q-1)^2}{1 + e^\frac{Q q}{T}}\,.
\label{eq:n2p_conversion}
\eal
where $g_A = 1.27$.
\end{widetext}
\end{appendix}

\end{document}